\documentclass[preprint,prb,floats,amsfonts,showpacs,eqsecnum,superscriptaddress,floatfix,multicol aps,showpacs]{revtex4}
\usepackage{amssymb}
\usepackage{epsfig}
\usepackage{amsmath}
\usepackage{psfrag}
\usepackage{color}
\usepackage{graphicx}
\usepackage{subfigure}
\usepackage{bm}

\usepackage{epsfig}

\newcommand{\be}{\begin{equation}}
\newcommand{\ee}{\end{equation}}
\newcommand{\bea}{\begin{eqnarray}}
\newcommand{\eea}{\end{eqnarray}}

\begin{document}
\draft
\title{The RKKY Interaction and the Nature of the Ground State of Double Dots in Parallel}

\author{M. Kulkarni$^{1,2}$ and R. M. Konik}
\affiliation{
Department of  Condensed Matter Physics and Material Science, Brookhaven National Laboratory, Upton, NY 11973-5000, USA\\
$^{2}$ Department of Physics and Astronomy, Stony Brook University, Stony Brook, NY 11794-3800, USA}
\date{\today}

\begin{abstract}
We argue through a combination of slave boson mean field
theory and the Bethe ansatz that the ground state of closely spaced double quantum dots in parallel coupled to a single effective channel are
Fermi liquids.  We do so by studying the dots' conductance, impurity entropy, and spin correlation.
In particularly we find that the zero temperature conductance is characterized by the Friedel sum
rule, a hallmark of Fermi liquid physics, and that the impurity entropy vanishes in the limit of zero temperature,
indicating the ground state is a singlet.  This conclusion is in opposition to a number of numerical renormalization group
studies.  We suggest a possible reason for the discrepancy.
\end{abstract}

\maketitle

\section{Introduction}

Quantum dots provide a means to realize strongly correlated physics in a controlled setting.
Because of the ability to adjust gate voltages which control both the tunnelling amplitudes between
the dots and the connecting leads and the dots' chemical potential,
quantum dots can be tuned to particular physical regimes.  One celebrated
example of said tuning was the first realization of Kondo physics in a single quantum dot,\cite{gold,cronenwett}
obtained by adjusting the chemical potential of the dot such that the dot was occupied by one electron.

More generally, engineered multi-dot systems offer the ability to realize more exotic forms of Kondo physics.
This was seen, for example, in the realization of the unstable
fixed point of two-channel overscreened Kondo physics in a multi-dot system.\cite{gold1}  There has thus been
considerable theoretical interest in double dots systems in different geometries, both in series (for example
Refs. \onlinecite{lee-2010,sela-2009,sela-2009qcp,Ireneusz-2010,malecki-sela-affleck,oguri}) and
in parallel (for example Refs. \onlinecite{short,long,zitko,
zitko-bonea-2006,hofstetter,ding,ding-ye,le-hur,vernek,ding-kim,luis-2006,luis-2008,sela-affleck-2009p,komnik}).
In this article we focus on the strongly correlated physics present on the latter geometry, in particular
when there is no direct tunneling between the dots and the dots are not capacitively coupled.
Such dot systems have been experimentally realized in numerous instances \cite{chen-2004,sigrist-2004,chen-review} (for other
realizations of double dots in parallel see Refs. \onlinecite{holleitner-2004,hubel-2008}).  Although
the dots are not directly coupled, they are coupled through an effective Ruderman-Kittel-Kasuya-Yosida
(RKKY) interaction.  It is aim of this work to explore the nature of the RKKY interaction in parallel double
quantum dots.

A straightforward application of the RKKY interaction as typically understood
would lead one to believe that in a closely spaced
double quantum dot with two electrons present, one on each dot, the RKKY interaction should lead to an
effective ferromagnetic coupling between the dots.  How should this coupling reveal itself in a transport
experiment, the typical probe of a quantum dot system?  If a ferromagnetic coupling is present, one expects
the electrons on the two dots to bind into a spin-1 impurity.  If the dots are coupled to a single
effective lead, we obtain, in effect, an underscreened Kondo effect.  As the temperature is lowered, the
single effective lead will partially screen the spin-1 impurity to an effectively uncoupled spin-1/2 impurity.
The ground state of the system will then be a non-Fermi liquid doublet.  In particular
at small temperatures and voltages, the
conductance through the dot will be characterized by logarithms.\cite{coleman}
This scenario has been put forth in a number of NRG studies \cite{zitko,zitko-bonea-2006,ding,ding-ye}
and is implicit in
a number slave boson studies
\cite{le-hur,vernek,ding-kim} of double dots in parallel.

We present contrary evidence here that this scenario is not in fact applicable at least for temperatures below the
Kondo temperature.  We argue that the ground state of closely spaced double
dots is instead a Fermi liquid singlet.  These findings are consistent with those of
Ref. \onlinecite{sela-2009}.  We do so using both the Bethe ansatz and slave boson mean field
theory.  It has been shown \cite{short,long} that under certain conditions
double dots in parallel admit an exact solution using the Bethe ansatz.  This exact solution, following
the approach introduced in Ref. \onlinecite{singledot}, can be exploited to compute transport properties.  In particular,
the zero temperature linear response conductance can be computed exactly.  Double quantum dots, however,
only admit an exact solution provided their parameters satisfy certain constraints.  To ensure that
our finding of Fermi liquid behaviour is not an artifact of these constraints, we also study the
parallel dot system using slave boson mean field theory.  This allows one to study
the sensitivity of our results to adding a second weakly coupled channel and to compute such quantities
as the spin-spin correlation function, an object not directly computable in the Bethe ansatz.

The paper is organized as follows.  In Section II we detail the double dot model that
we are interested in studying together
with the approaches (Bethe ansatz and slave boson mean field theory) that we employ in studying this
system.  In Section III we present results on the linear response conductance through the dots both at
zero temperature and finite temperature.  We show the zero temperature conductance obeys the Friedel
sum rule, a hall mark of Fermi liquid physics.  We also study the
impurity entropy at finite temperature
showing that it vanishes in the zero temperature limit indicating the presence of a singlet.
Finally in this section, we present results (using slave boson mean field theory alone) of the spin-spin
correlation function.
Lastly, in Section IV, we discuss the implications of our results and suggest
a way they can be reconciled with the conflicting NRG studies.

\section{Model Studied}

We study a set of two dots arranged in parallel with two leads.  The Hamiltonian for this system is given
by
\begin{eqnarray}\label{eIIi}
{\cal H} &=& -i\sum_{l\sigma}\int^\infty_{-\infty} dx c^\dagger_{l\sigma}\partial_xc_{l\sigma}
+ \sum_{\sigma\alpha}V_{l\alpha}(c^\dagger_{l\sigma\alpha}d_{\sigma\alpha} + {\rm h.c}) \cr\cr
&& + \sum_{\sigma\alpha}\epsilon_{d\alpha}n_{\sigma\alpha} + \sum_{\alpha} U_{\alpha}
n_{\uparrow\alpha}n_{\downarrow\alpha}.
\end{eqnarray}
The $c_{l\sigma}$ specify electrons with spin $\sigma$ living on the two leads, $l=1,2$.  The $d_{\alpha\sigma}$ specify
electrons found on the two dots $\alpha=1,2$.  Electrons can hop from the leads to dots with tunneling strength
$V_{l\alpha}$.  The strength of the Coulomb repulsion on the two dots is given by $U_\alpha$.  We suppose that
there is no interdot Coulomb repulsion and that tunneling between the two dots is negligible.
A schematic of this Hamiltonian for the two dots is given in Fig. 1.

\begin{figure}
\begin{center}
\includegraphics[scale=0.6,angle=-90]{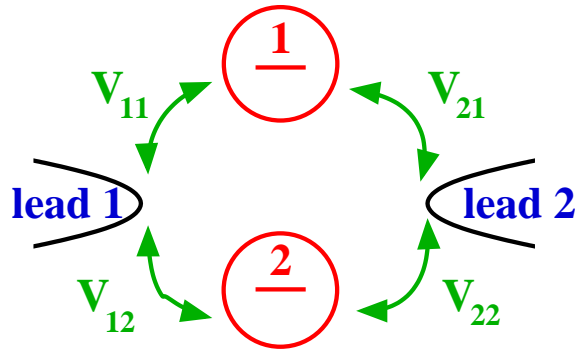}
\caption{A schematic of the double dot system.}
\end{center}
\end{figure}

\subsection{Bethe Ansatz}

The above Hamiltonian can be solved exactly via Bethe ansatz under certain conditions.  The set of constraints
that we will be particularly interested in are as follows:
\begin{eqnarray}\label{eIIii}
V_{1\alpha}/V_{2\alpha} &=& V_{1\alpha'}/V_{2\alpha'};\cr\cr
U_{\alpha} \Gamma_\alpha &=& U_{\alpha'}\Gamma_{\alpha'};\cr\cr
U_{\alpha} + 2\epsilon_{d\alpha} &=& U_{\alpha'} + 2\epsilon_{d\alpha'}.
\end{eqnarray}
The first of these conditions results in only a single effective channel coupling to the two leads.
This occurs automatically when the dot hoppings are chosen to be symmetric and so is commonly found
in the literature.\cite{le-hur,c1,zitko,zitko-bonea-2006,ding-kim,luis-2006,luis-2008}
The second condition tells us that with $U_{1,2}$ fixed,
$\epsilon_{d1}-\epsilon_{d2}$, is also fixed.  We thus must move $\epsilon_{d1,2}$ in unison in order
to maintain integrability.
The final condition tells us that $\sqrt{U_i\Gamma_i}$, the bare scale governing charge fluctuations
on the dots must be the same on both dots.

To solve this Hamiltonian we implement a map to even and odd channels,
$c_{e/o}=(V_{1/2,\alpha}c_1 \pm V_{2/1,\alpha}c_2)/\sqrt{2\Gamma_\alpha}$ where $\Gamma_\alpha = (V^2_{1\alpha}+V^2_{2\alpha})/2$.
Under the map, the Hamiltonian factorizes into even and odd sectors:
\begin{eqnarray}\label{eIIiii}
{\cal H}_e &=& -i\sum_{l\sigma}\int^\infty_{-\infty} dx ~c^\dagger_{e\sigma}\partial_x c_{e\sigma}
+ \sum_{\sigma\alpha}\sqrt{2\Gamma_{\alpha}}(c^\dagger_{e\sigma\alpha}d_{\sigma\alpha} + {\rm h.c}) \cr\cr
&& + \sum_{\sigma\alpha}\epsilon_{d\alpha}n_{\sigma\alpha} +
\sum_{\alpha} U_{\alpha} n_{\uparrow\alpha}n_{\downarrow\alpha};\cr
{\cal H}_o &=& -i\sum_{l\sigma}\int^\infty_{-\infty} dx ~c^\dagger_{o\sigma}\partial_x c_{o\sigma},
\end{eqnarray}
where, as can be seen, the odd sector decouples from the double dot.  Using the Bethe ansatz \cite{short,long}
one can construct N-particle wave functions in the non-trivial even sector.  These wavefunctions are characterized
by N-momenta $\{q_i\}^N_{i=1}$ and M quantum numbers $\{\lambda_\alpha\}^M_{\alpha = 1}$.  The number of $\lambda_\alpha$'s
mark the spin quantum number of the wave function: $S_z=N-2M$.  Together the $\lambda_\alpha$'s and $q_i$'s
satisfy the following set of constraints:
\begin{eqnarray}\label{eIIiv}
e^{iq_jL+i\delta(q_j)} &=& \prod^M_{\alpha = 1}\frac{g(q_j)-\lambda_\alpha+i/2}{g(q_j)-\lambda_\alpha-i/2};\cr\cr
\prod^N_{j=1}\frac{\lambda_\alpha-g(q_j)+i/2}{\lambda_\alpha-g(q_j)-i/2}
&=& -\prod^M_{\beta=1}\frac{\lambda_\alpha-\lambda_\beta+i}{\lambda_\alpha-\lambda_\beta-i},
\end{eqnarray}
where $g(q) = \frac{(p-\epsilon_{d\alpha}-U_\alpha/2)^2}{2\Gamma_\alpha U_\alpha}$.
These equations are identical to the set of constraints for a single dot \cite{wie,wie1} but for the form of the scattering
phase $\delta(q)$:
\begin{equation}
\delta (q) = -2\tan^{-1}(\sum_\alpha \frac{\Gamma_\alpha}{q - \epsilon_{d\alpha}} ).
\end{equation}

We will focus in this paper on computing transport properties in linear response.
At zero temperature we can use the Bethe ansatz to access such transport quantities exactly.\cite{singledot,long}  We will
also use the Bethe ansatz to obtain an excellent quantitatively accurate prediction (in comparison with NRG)
for the finite temperature linear response conductance (see Refs. \onlinecite{singledot,long} for the Bethe
ansatz computation of the finite temperature linear response conductance for a single dot and for the
comparable NRG, Ref. \onlinecite{costi}).

The Bethe ansatz can be exploited
to develop certain approximations that allow one
to compute certain non-equilibrium quantities, in particular, the out-of-equilibrium conductance \cite{singledot}
and the noise \cite{noise} in the presence of a magnetic field.  In order to obtain at least qualitatively accuracy, the presence
of a magnetic field is a necessity.  With a magnetic field the Bethe ansatz for a single dot correctly predicts such features as
the positioning of the peak in the differential magnetoconductance \cite{noise} as say measured in
carbon nanotube quantum dots.\cite{gold_nano}  In the absence of a magnetic field, the Bethe ansatz inspired
approximation breaks down however.\cite{noise}  We will, again, not consider the double dots out-of-equilibrium
in this work.

\subsection{Slave Boson Mean Field Theory}

We also study the Hamiltonian (\ref{eIIi}) using a slave-boson mean field
theory, a well-known technique, applicable at sufficiently low temperatures.\cite{hewson}
The starting point is the same Hamiltonian (\ref{eIIi}) and we will study
here the $U_{\alpha}=\infty$ case. The constraint of preventing
double-occupancy on the dots is fulfilled by introducing two Lagrange
multipliers
$\lambda_{1}$ and $\lambda_{2}$. The slave boson formalism consists
of writing the impurity fermionic operator on each dot as a combination
of a pseudofermion and a boson operator: $d_{\sigma\alpha}=b_{\alpha}^{\dagger}f_{\sigma\alpha}$.
Here $f_{\sigma\alpha}$ is the pseudofermion which annihilates one
``occupied state'' on dot $\alpha$ and $b_{\alpha}^{\dagger}$
is a bosonic operator which creates an empty state on dot $\alpha$.
The mean field approximation consists of replacing the bosonic
operator by its expectation value:
$b_{\alpha}^{\dagger}\rightarrow\left\langle b_{\alpha}^{\dagger}\right\rangle =r_{\alpha}$.
Thus $r_{\alpha}$ and $\lambda_{\alpha}$ together form four parameters
which need to be determined using mean field equations. Under the
slave boson formalism combined with the mean field approximation,
Eq. (\ref{eIIi}) reads
\begin{eqnarray}\label{eq:SBMFT_H}
H_{SBMFT} & = & -i\sum_{l\sigma}\int_{-\infty}^{+\infty}dxc_{l\sigma}^{\dagger}\partial_{x}c_{l\sigma}+\sum_{l\alpha\sigma}\tilde{V}_{l\alpha}\left(c_{l\sigma}^{\dagger}f_{\sigma\alpha}+\mbox{h.c}\right)\nonumber \\
 & + & \sum_{\sigma\alpha}\tilde{\epsilon}_{d\alpha}f_{\sigma m}^{\dagger}f_{\sigma m}+i\sum_{\alpha}\lambda_{\alpha}(r_{\alpha}^{2}-1)
\end{eqnarray}
with $\tilde{V}_{l\alpha}=r_{\alpha}V_{l\alpha}$ and
$\tilde{\epsilon}_{d\alpha}=\epsilon_{d\alpha}+i\lambda_{\alpha}$.
The mean field equations are the constraints for the dot $\alpha=1,2$,
\begin{equation}\label{constraint}
\sum_{\sigma}<f_{\alpha\sigma}^{\dagger}(t)f_{\alpha\sigma}(t)>+r_{\alpha}^{2}=1,
\end{equation}
and the equation of motion (EOM) of the boson fields,
\begin{equation}\label{eom_bose}
{\rm Re}\left[\sum_{l,k,\sigma}\tilde{V}_{l\alpha}^{*}\left\langle c_{kl\sigma}^{\dagger}(t)f_{\sigma\alpha}(t)\right\rangle
\right]+i\lambda_{\alpha}r_{\alpha}^{2}=0.
\end{equation}
The above equations can be understood as arising from the conditions
\begin{eqnarray}
\frac{\partial\left\langle H_{SBMFT}\right\rangle }{\partial\lambda_{\alpha}} &=& 0; \cr\cr
\frac{\partial\left\langle H_{SBMFT}\right\rangle }{\partial r_{\alpha}} &=& 0.
\end{eqnarray}
Thus the reality condition on Eqn. 2.8 arises from the reality of the hopping
term in the Hamiltonian.\cite{newns}
For any given set of bare parameters ($\epsilon_{d\alpha\sigma},V_{l\alpha}$)
we can compute the renormalized energy ($\tilde{\epsilon}_{d\alpha\sigma}$)
and renormalized hybridization ($\tilde{V}_{l\alpha}$) by solving
the four equations, Eqns. \ref{constraint} and \ref{eom_bose}.
While these results
are mean field, they allow one to
span a wide parameter space not constrained by the requirements
of integrability. For instance, we study asymmetrically coupled
dots where two channels couple to the dot, a case not solvable by the Bethe
Ansatz. SBMFT allows one also, unlike the Bethe ansatz, to readily
study such quantities as the spin-spin correlation function.

\section{Results}

In this section we present a number of measures as computed using both the Bethe ansatz and slave boson mean
field theory
that are indicative of the ground state of the double dot plus lead
system.  We will argue that these are consistent with the ground state of the dot being a singlet
state, not a doublet.

\begin{figure}
\begin{center}
\includegraphics[scale=0.6,angle=-90]{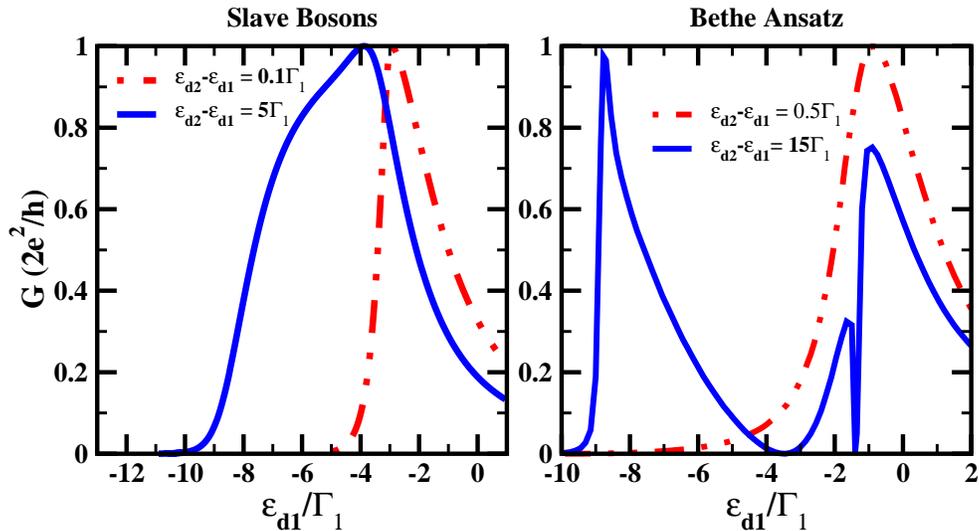}
\caption{The zero temperature conductance of a symmetrically coupled double dot computed using slave boson mean field theory and the Bethe ansatz.}
\end{center}
\end{figure}

\subsection{Zero Temperature Conductance}

The first measure we examine is the zero temperature linear response conductance, $G$.
For the BA, $G$ is computed as is discussed in great detail in Refs. \onlinecite{singledot} and \onlinecite{long}.
For the SBMFT, $G$ is computed by first solving for the variational parameters, $r_\alpha$, and $\tilde\epsilon_{d\alpha}$,
$\alpha = 1,2$, and then determining the corresponding transmission amplitude via solving
a one-particle Schr\"odinger equation.
This procedure is detailed in Appendix B.

If the double dot
is in a singlet state, we expect $G$ to vanish as $\epsilon_{d1,2}$ are lowered, moving the dot into the
Kondo regime.  This is consistent with understanding the singlet state as a Fermi liquid state.  If a
Fermi liquid, $G$ will obey the Friedel sum rule:\cite{langreth}
\begin{equation}
G = \sum_{\sigma = \uparrow,\downarrow}\frac{e^2}{h}\sin^2(\pi n_{d\sigma}),
\end{equation}
where $n_{d\sigma}$ is the number of electrons of spin species $\sigma$ displaced by the dot.
Deep in the Kondo regime, there will be two electrons sitting on the two dots, one of each spin
species, i.e. $n_{d\sigma}=1$, and so $G$ correspondingly vanishes.

\begin{figure}
\begin{center}
\includegraphics[scale=0.6,angle=-90]{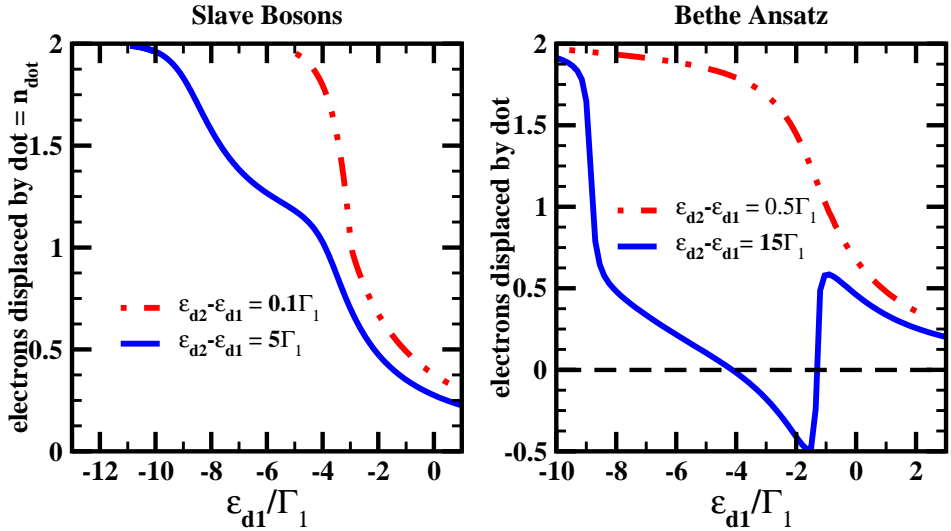}
\caption{The number of electrons displaced by the dots, $n_d$ of a symmetrically
coupled double dot computed both using SBMFT and the Bethe ansatz.  In the case
of SBMFT, $n_d$ is simply the dot occupation, $n_{dot}=\sum_{i\sigma}\langle d^\dagger_{i\sigma}d_{i\sigma}\rangle$.
In the case of the Bethe
ansatz, the quantity plotted is equal to the dot occupation {\it plus} the $1/L$ correction
to the electron density in the leads induced by coupling the dots to the leads.}
\end{center}
\end{figure}

Plotted in Fig. 2 is the zero temperature conductance as computed with BA and SBMFT as a function
of $\epsilon_{d1}$ and with $\epsilon_{d1}-\epsilon_{d2}$ fixed.  For each computational methodology we present
results for both $\epsilon_{d1}-\epsilon_{d2} \gg \Gamma_{1,2}$ and $\epsilon_{d1}-\epsilon_{d2} \ll \Gamma_{1,2}$.
We see that in all cases that as $\epsilon_{d1,2}$ is lowered, the conductance vanishes.  We do note
however that the overall structure of the conductance differs as computed between the BA and SBMFT, that
is to say, the SBMFT fails in general to describe the correct behaviour.  In particular it fails
to describe the intermediate valence regime when the distance separating the chemical potential
of the two dots is large, i.e $\epsilon_{d1}-\epsilon_{d2} \gg \Gamma_{1,2}$ .  In the intermediate valence regime
the conductance of the double dot as computed with the BA undergoes rapid changes, a consequence
of interference between electrons tunneling off and on the dot (see Ref. \onlinecite{long}).  This is not mirrored
in the SBMFT computations which remain comparatively structureless.

The failure of SBMFT to accurately capture the physics in the intermediate valence regime
is seen in a related quantity, the number of electrons displaced by the dot. $n_d$ is the sum of two terms:
\begin{equation}\label{nd}
n_d = \sum_{\sigma i}\langle d^\dagger_{\sigma i}d_{\sigma i}\rangle +
\sum_{\sigma l}\int dx \bigg[\langle c^{\dagger}_{\sigma l}c_{\sigma l}\rangle - \rho_{\sigma_{\rm bulk}}\bigg].
\end{equation}
The first term is simply the occupancy of the double dots while the second term measures the deviation
of electron density in the leads due to coupling the dots and the leads.  In SBMFT this term is zero
due to the mean field nature of its approximation.  However in BA this term is non-zero.  While we cannot
compute it directly, the BA gives us the ability to compute $n_d$.  And as plotted on the r.h.s. of Fig. 3,
we see that $n_d$ can be negative.  As $\sum_{\sigma i}\langle d^\dagger_{\sigma i}d_{\sigma i}\rangle$
is manifestly a positive
quantity, we know that as computed by the BA, the second term in Eq.(\ref{nd}) is non-zero and in fact
is negative (at least in the case $\epsilon_{d1}-\epsilon_{d2} \gg \Gamma_{1,2}$).
From Fig. 3 we see however that $n_d$ for both SBMFT and BA tends to two as the system enters
the Kondo regime (where two electrons sit on the two dots).

\begin{figure}
\begin{center}
\includegraphics[scale=0.4,angle=-90]{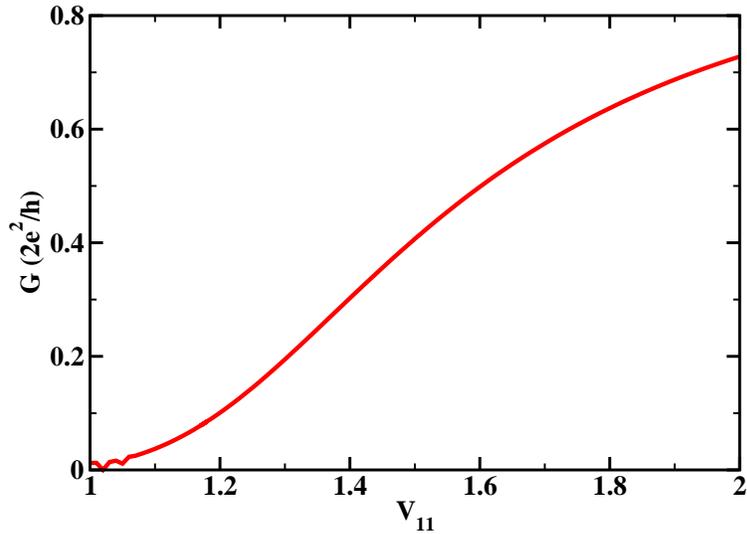}
\caption{The zero temperature conductance of asymmetrically coupled double dot computed using slave boson mean field theory. The conductance is plotted as a function of $V_{11}$.  The remaining dot-lead hopping strengths are all
set to 1 while $\epsilon_{d1}=-4.7$ and $\epsilon_{d2}=-4.6$.}
\end{center}
\end{figure}

One advantage the SBMFT does offer over the BA is that it allows us to compute transport quantities
beyond the integrable parameter regime delineated in Eq. (\ref{eIIii}).  It was argued in Ref.\onlinecite{long}
that small deviations away from this parameter space should not qualitatively change transport properties.
In Fig. 4 we test this in the Kondo regime (where we expect SBMFT to be at its most accurate)
computing the conductance while
adjusting the dot-lead hopping parameters in such a way that we move
from a case where only one effective channel couples to the quantum dot (i.e. $V_{ij}=1$) to
a case where two channels couple to the dot ($V_{11}>1$ and $V_{12}=V_{21}=V_{22}=1$).  We see
when the second channel is only weakly coupled to the dot, the conductance remains near its one-channel
value, i.e. $G=0e^2/h$.  Only once $V_{11}$ appreciably deviates from 1 do we see a corresponding deviation
in $G$.  We note that this continuous behaviour is also consistent with the one-channel dot-lead ground state
being a singlet.  If it were instead a doublet, coupling a second channel into the system would lead to
a discontinuous change as the second channel, no matter how weakly coupled, would immediately screen
the free effective spin-1/2.

\begin{figure}
\begin{center}
\includegraphics[scale=0.6,angle=-90]{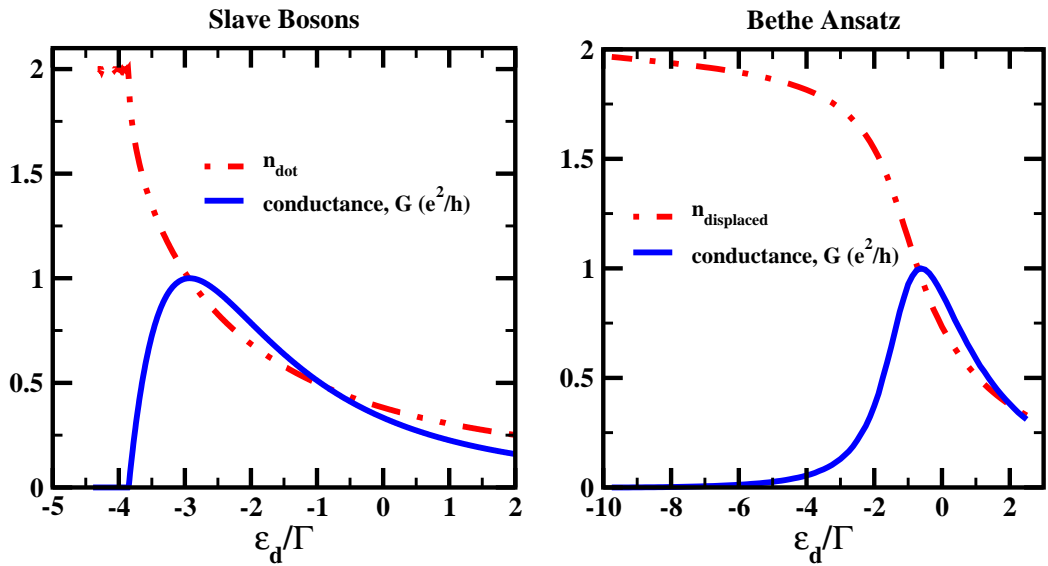}
\caption{The conductance and the number of displaced electrons as a function of $\epsilon_{d1}$(=$\epsilon_{d2}$)
as computed using SBMFT and the Bethe ansatz.}
\end{center}
\end{figure}

Finally in this section we consider the behaviour of the conductance and displaced electrons when
$\epsilon_{d1}=\epsilon_{d2}$.  We see from Fig. 5 that the same qualitative behaviour in both quantities
is found using the Bethe ansatz and using SBMFT.  Namely, the displaced electron number $n_d$ tends
to 2 while the conductance $G$ tends to zero as $\epsilon_{d1}=\epsilon_{d2}$ goes to zero.  While
these measures are the same in the two computational methods, there is a quantity that sharply
distinguishes the two (and so shows a failure of SBMFT even in the Kondo regime at zero temperature).
This quantity is the low lying density of states on the dots, $\rho_d(\omega)$.  For the case
of $\epsilon_{d1}=\epsilon_{d2}$, the BA shows that $\rho_d(\omega)$ for $\omega$ on the order
of the Kondo temperature, $T_K$ vanishes.\cite{long}  However the SBMFT shows that at this energy
scale there exists non-negligible spectral weight.  In Fig. 6, we plot $\rho_d(\omega)$ as defined by
$$
\rho_d (\omega ) = \sum_{i\sigma}{\rm Im}\langle d_{i\sigma}^\dagger d_{i\sigma}\rangle_{retarded}.
$$
The agreement on the qualitative features of $n_d$ and $G$ between the two methodologies is then a coincidence
(to a degree).
In both cases the ground state is Fermi liquid like and so $G$ follows the Friedel sum rule which necessarily
mandates
that the conductance vanish with two electrons on the two dots.  But the nature of the Fermi liquid in each
case as predicted by the methodologies is much different.  SBMFT predicts the scale of the low lying excitations
is $T_K$ while the BA finds that for the special case of $\epsilon_{d1}=\epsilon_{d2}$ (and only
for this case\cite{long}) that the fundamental energy
scale corresponds to the bare energies scales in the problem, i.e. $U$ and $\Gamma$.

\begin{figure}
\begin{center}
\includegraphics[scale=0.3,angle=-90]{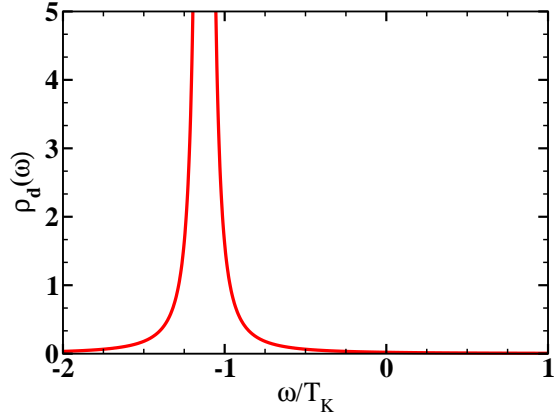}
\caption{A plot of the low energy density of states, $\rho (\omega)$ for $\epsilon_{d1}=\epsilon_{d2}$
in the Kondo regime as computed using SBMFT.  As we are argue in the text,
this is an artifact of SBMFT (the BA shows that in this case $\rho_d(\omega )$ vanishes\cite{long}).}
\end{center}
\end{figure}

\subsection{Finite Temperature Conductance}

We now consider the finite temperature conductance.  Plotted in Fig. 7 are traces for $G(T)$ for double
dots with both $|\epsilon_{d1}-\epsilon_{d2}| \gg \Gamma_{1,2}$ and
$|\epsilon_{d1}-\epsilon_{d2}| \ll \Gamma_{1,2}$ as computed with both the Bethe ansatz as well as SBMFT.
For a Fermi liquid ground state we expect
that at low temperatures the conductance deviate from its zero temperature value by $T^2$ and we
see that behaviour in both cases.  From Fig. 7 we see that in both treatments,
the finite temperature conductance for the double
dots initial rises with increasing temperature to an appreciable fraction of the unitary maximum and thereafter
decreases in an uniform manner  (regardless of the bare level separation).  (For SBMFT we have defined the Kondo
temperature as where this peak in $G(T)$ occurs while for the BA the Kondo temperature we employ the analytic
expression for $T_K$ of the double dots derived in
Ref. \onlinecite{long}.)  We however also see
that there are qualitative differences between SBMFT and the Bethe ansatz.  The peak in the conductance
computed using SBMFT peaks at a value far closer to the unitary maximum than does the Bethe ansatz.  And we
also see that the conductance as computed in the SBMFT drops off far more rapidly than it does in the
BA (particularly at large level separation).  We however believe this is unphysical and akin to the
pathologies that SBMFT is known to exhibit at higher temperatures
and energy scales.\cite{newns,aligia,kang,nunes,schlottmann,kroha,wolfle}

\begin{figure}
\begin{center}
\includegraphics[scale=0.6,angle=-90,trim = 0mm 0mm 0mm 0mm,clip]{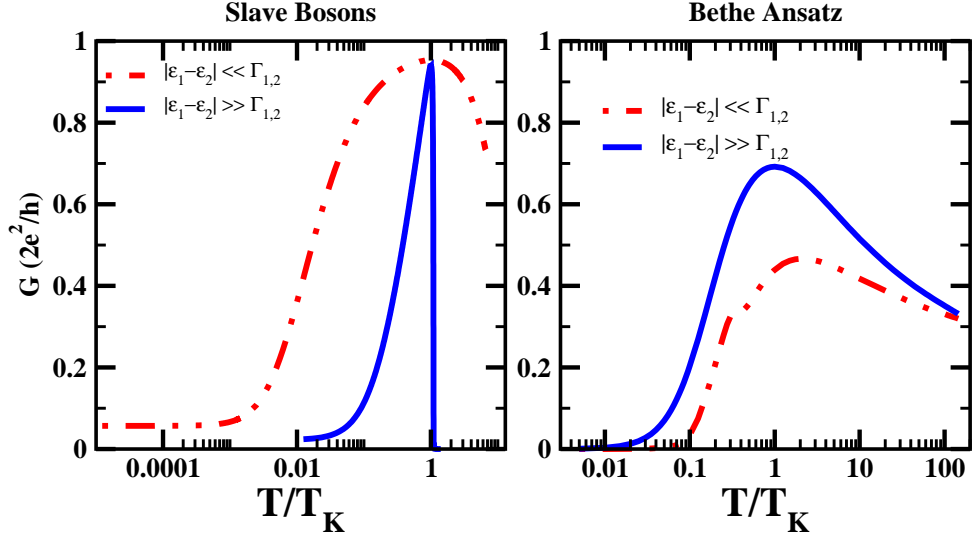}
\caption{The linear response conductance as a function of temperature
of a symmetrically coupled double dot computed using both slave boson mean field theory and the Bethe ansatz.}
\end{center}
\end{figure}

As was demonstrated in Ref. \onlinecite{long}, the conductance at finite but small $T$ is quadratic in $T$ while at large $T$
the conductance is logarithmic (going as $1/\log^2(T/T_K)$).  The peak in conductance at finite $T$
is then a result of the conductance vanishing in the low and high temperature limits.  These conductance profiles are
similar to the those predicted in Ref. \onlinecite{pustilnik} for multi-dots coupled to {\it two} electron channels.
However the physics
there is much different: the non-monotonicity in $G(T)$ predicted in Ref. \onlinecite{pustilnik} is due to
the presence of the two channels and because they couple to the dots with different strengths, they screen
a $S>1/2$ state in stages.

\subsection{Spin-Spin Correlation Function}

We present the static spin-spin correlation function as a function of $\epsilon_{d1}$ and as computed using SBMFT in Fig. 8.
With two electrons on the dot
the value of $\langle S_1\cdot S_2\rangle$ can vary between $-3/4$ (if these two electrons are bound in a singlet state)
to $1/4$ (if the two electrons find themselves in a triplet state).
We see in Fig. 8 that generically the value of the correlation function in the Kondo regime (for the relevant values
of $\epsilon_{d1}$ see Fig. 2) that $\langle S_1\cdot S_2\rangle$ tends to 0.  This however should not necessarily be interpreted
as the dots being closer to a triplet state than a singlet state.  In determining the overall state of the system,
$\langle S_1\cdot S_2\rangle$, is not necessarily a good measure.  We can see this by considering a simple toy example.

Imagine a system of four spins, two associated with the dots, $|\uparrow\rangle_{d1}$ and $|\uparrow\rangle_{d2}$, and
two associated with leads $|\uparrow\rangle_{l1}$ and $|\uparrow\rangle_{l2}$.  And first suppose the system is in a singlet state.
Two ways this singlet state can be formed are
\begin{eqnarray}
|{\rm singlet~ 1}\rangle &=& \frac{1}{2} (|\uparrow\rangle_{d1}|\downarrow\rangle_{d2}-|\downarrow\rangle_{d1}|\uparrow\rangle_{d2})
\otimes (|\uparrow\rangle_{l1}|\downarrow\rangle_{l2}-|\downarrow\rangle_{l1}|\uparrow\rangle_{l2});\cr\cr
|{\rm singlet~ 2}\rangle &=& \frac{1}{2} (|\uparrow\rangle_{l1}|\downarrow\rangle_{d1}-|\downarrow\rangle_{l1}|\uparrow\rangle_{d1}) \otimes
(|\uparrow\rangle_{l2}|\downarrow\rangle_{d2}-|\downarrow\rangle_{l2}|\uparrow\rangle_{d2}).
\end{eqnarray}
We see that the expectation value, $\langle S_1\cdot S_2\rangle$, of these two states is considerably different:
\begin{eqnarray}
\langle {\rm singlet~ 1}|S_1\cdot S_2|{\rm singlet ~1}\rangle &=& -3/4\cr\cr
\langle {\rm singlet~ 2}|S_1\cdot S_2|{\rm singlet ~2}\rangle &=& 0.
\end{eqnarray}

\begin{figure}
\begin{center}
\includegraphics[scale=0.4,angle=-90]{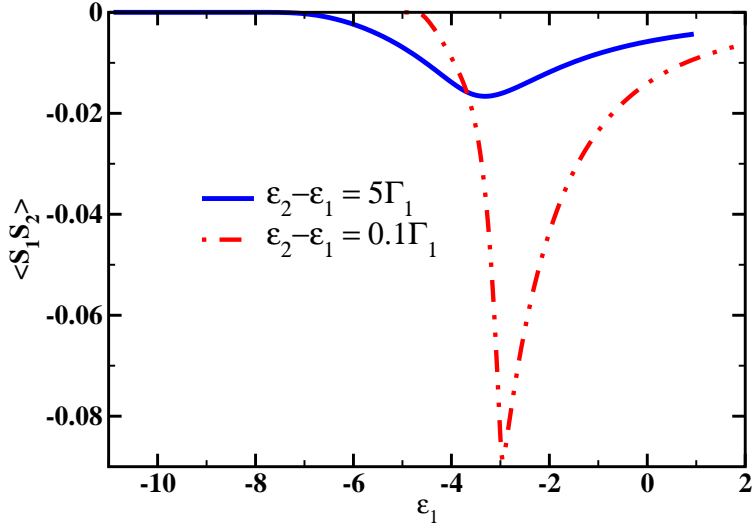}
\caption{The spin-spin correlation function of symmetrically coupled double dot computed using slave boson mean field theory.}
\end{center}
\end{figure}
Now suppose the system is in a triplet state and suppose its $S_z$ projection is 1.  Again there are two
inequivalent ways this state can be formed:
\begin{eqnarray}
|{\rm triplet~ 1}\rangle &=& \frac{1}{\sqrt{2}} |\uparrow\rangle_{d1}|\uparrow\rangle_{d2}
\otimes (|\uparrow\rangle_{l1}|\downarrow\rangle_{l2}-|\downarrow\rangle_{l1}|\uparrow\rangle_{l2});\cr\cr
|{\rm triplet~ 2}\rangle &=& \frac{1}{\sqrt{2}} |\uparrow\rangle_{l1}|\uparrow\rangle_{d1} \otimes
(|\uparrow\rangle_{l2}|\downarrow\rangle_{d2}-|\downarrow\rangle_{l2}|\uparrow\rangle_{d2}).
\end{eqnarray}
The expectation of these two values is
\begin{eqnarray}
\langle {\rm triplet~ 1}|S_1\cdot S_2|{\rm triplet ~1}\rangle &=& 1/4\cr\cr
\langle {\rm triplet~ 2}|S_1\cdot S_2|{\rm triplet ~2}\rangle &=& 0.
\end{eqnarray}
We thus see that when the system's state is such that $\langle S_1\cdot S_2\rangle = 0$, it can be either a singlet or
a triplet equally.  We thus end with a more reliable measure of the dot's internal degrees of freedom: the impurity entropy.

\subsection{Impurity Entropy}

The final set of computations we present in this section are of the impurity entropy of the double
dots.  In Fig. 9 we plot results coming from SBMFT and the BA for both large and small level separation.
(The derivation of the impurity entropy in the context of the BA is found in Appendix A.)
We see that in all cases the impurity entropy vanishes as $T\rightarrow 0$.  This then implies
the ground state of the double dot system is a singlet.  If it were a triplet state, the $T\rightarrow 0$ limit would
lead to $S_{imp} = \log (2)$.

\begin{figure}
\begin{center}
\includegraphics[scale=0.6,angle=-90,trim = 0mm 0mm 0mm .1mm,clip]{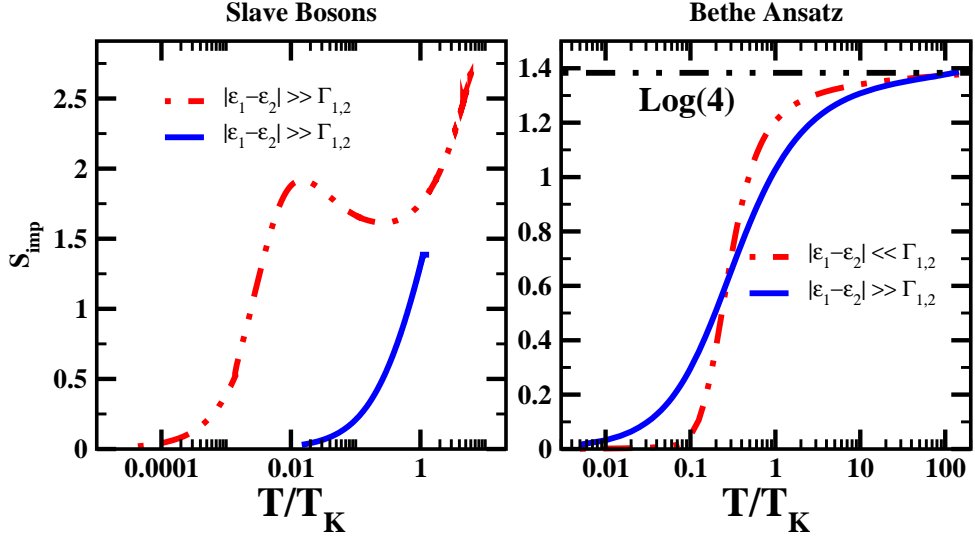}
\caption{The impurity entropy as a function of temperature
of a symmetrically coupled double dot computed using slave boson mean field theory and the Bethe ansatz.}
\end{center}
\end{figure}

\section{Discussion and Conclusions}

We have presented a number of arguments that the ground state of a double dot near the particle-hole symmetric point
(i.e. when there are nearly two electrons on the dots) is in a singlet Fermi liquid state.  In particular
we have shown that the conductance in this limit vanishes, in accordance with the Friedel sum rule and that
the impurity entropy also vanishes,
in agreement with the ground state being a singlet.

These conclusions however disagree with a number of NRG studies.  In Refs. \onlinecite{zitko}
and \onlinecite{zitko-bonea-2006} it is
found that a double quantum dot (with the dots closely spaced) carrying two electrons is in a
spin-triplet state, has a conductance
corresponding to the unitary maximum, and is correspondingly a non-Fermi liquid.  Similar conclusions are reached in
Refs. \onlinecite{ding,ding-ye}.  The basic rational invoked for observing this physics
is that with dots closely spaced, a ferromagnetic
RKKY interaction is present which binds the two electrons on the dot into a spin triplet.
Consequently the ground state of the double
dot is that of an underscreened spin-1 Kondo impurity.

We offer a possible reason for the discrepancy that we find
with these studies.  The Anderson Hamiltonian typically considered in these
studies is of the form:
\begin{eqnarray}
H &=& H_{lead} + H_{dot} + H_{lead-dot}; \cr\cr
H_{lead} &=& D\int^{1}_{-1}k dk a^\dagger_{k\sigma}a_{k\sigma};\crcr
H_{dot} &=& \sum_{i=1,2;\sigma}\epsilon_{di}d^\dagger_{i\sigma}d_{i\sigma} + \sum_iU_in_{i\uparrow}n_{i\downarrow};\cr\cr
H_{dot-lead} &=& D^{1/2}\sum_{i\sigma}\int^{1}_{-1} dk ~V_i(d^\dagger_{i\sigma}a_{k\sigma}+a^\dagger_{k\sigma}d_{i\sigma})
\end{eqnarray}
where we are using the conventions of Ref. \onlinecite{krishna} in writing down the NRG Hamiltonian.
Before implementing the NRG algorithm,
one adopts a logarithmic basis
for the lead electrons,
\begin{eqnarray}
a_{k\sigma} &=& \sum_{np}a_{np\sigma}\psi^+_{np}(k) + b_{np\sigma}\psi^{-}_{np}(k),
\end{eqnarray}
where
\begin{equation}
\psi^{\pm}_{np}(k) =
\begin{cases}
\frac{\Lambda^{n/2}}{(1-\Lambda^{-1})^{1/2}}e^{\pm i\omega_n pk} & \text{if $\Lambda^{-(n+1)}< \pm k < \Lambda^{-n}$,} \\
0 & \text{\rm if~k~is~not~within~the~above~interval,}
\end{cases}
\end{equation}
and $w_n$ is given by
\begin{equation}
w_n = \frac{2\pi\Lambda^n}{1-\Lambda^{-1}}.
\end{equation}
Here $\Lambda$ is a parameter less than one.
This change of basis transforms
$H_{lead}$ and $H_{lead-dot}$ into
\begin{eqnarray}
H_{lead} &=&
\frac{D}{2}(1+\Lambda^{-1})\sum_{np}\Lambda^{-n}(a^\dagger_{np\sigma}a_{np\sigma} - b^\dagger_{np\sigma}b_{np\sigma})\cr\cr
&& + \frac{D(1-\Lambda^{-1})}{2\pi i}\sum_n\sum_{p\neq p'}\exp(\frac{2\pi i (p'-p)}{1-\Lambda^{-1}})
(a^\dagger_{np\sigma}a_{np'\sigma} - b^\dagger_{np\sigma}b_{np'\sigma})\frac{\Lambda^{-n}}{p'-p};\cr\cr
H_{lead-dot} &=& D^{1/2}(1-\Lambda^{-1})^{1/2}\sum_{in\sigma} \Lambda^{-n/2}V_i ((a^\dagger_{n0\sigma}+b^\dagger_{n0\sigma})d_i + {\rm h.c.}).
\end{eqnarray}
Typically the approximation that is now made in most NRG treatments (and seems to have been made in the above
references) is that the $p\neq0$ modes of the logarithmic basis are dropped, not least because these modes
do not directly couple to dot.  In the single dot case, where
this approximation was first made,\cite{wilson,krishna} this was
found to be a reasonable approximation.  However in
the double dot case it is not {\it a priori} obvious that this is the case.  In particular
if one finds an underscreened Kondo effect one might ask whether that additional modes ($p\neq 0$) might
serve to provide additional screening channels.

Let us then consider the NRG Hamiltonian that would arise if both the $p=0$ and $p=\pm 1$ modes were kept.
$H_{lead}$ can then be trivially diagonalized using the combination
\begin{eqnarray*}
r_{0\sigma} & = & \frac{1}{3}\left(2a_{n1\sigma}-e^{\theta}a_{n0\sigma}+2e^{2\theta}a_{n-1\sigma}\right);\cr\cr
r_{\pm\sigma} & = & \frac{1}{3\sqrt{10}}\left[5a_{n1\sigma}+(2\mp6i)e^{\theta}a_{n0\sigma}-(4\pm3i)e^{2\theta}a_{n-1\sigma}\right],
\end{eqnarray*}
with $\theta=-\frac{2\pi i}{1-\Lambda^{-1}}$. Note that corresponding
transformation $\{s\leftrightarrow b\}$ is omitted for brevity. This
yields
\begin{eqnarray*}
H_{lead} & = & D\sum_{n\sigma}\Lambda^{-n}\Biggl\{\frac{1+\Lambda^{-1}}{2}r_{0\sigma}^{\dagger}r_{0\sigma}+\frac{2\pi(1+\Lambda)+3(1-\Lambda)}{4\pi\Lambda}r_{+\sigma}^{\dagger}r_{+\sigma}\\
 &  & +\frac{2\pi(1+\Lambda)-3(1-\Lambda)}{4\pi\Lambda}r_{-\sigma}^{\dagger}r_{-\sigma}-\{r\rightarrow s\}\Biggr\}\end{eqnarray*}
The corresponding transformation of dot-lead Hamiltonian is
\begin{equation}
H_{dot-lead}=\sum_{in\sigma}V_{ni}\left[\sqrt{\frac{2}{5}}\left(\frac{1}{3}-i\right)r_{+\sigma}^{\dagger}d_{i\sigma}+\sqrt{\frac{2}{5}}\left(\frac{1}{3}+i\right)r_{-\sigma}^{\dagger}d_{i\sigma}-\frac{1}{3}r_{0\sigma}^{\dagger}d_{i\sigma}+\{r\rightarrow s\}+h.c\right],
\end{equation}
with
$V_{ni}=D^{\frac{1}{2}}(1-\Lambda^{-1})^{\frac{1}{2}}\Lambda^{-\frac{n}{2}}e^{\theta}V_{i}$.

We see upon this diagonalization
that three channels of electrons, $r_{n\{0,\pm\},\sigma}$ and $s_{n\{0,\pm\}\sigma}$, couple to the dot.  And because of the nature
of the logarithmic basis, we see that $+$, $-$ and $0$ variables can be arbitrarily
close to the Fermi surface (due to the presence of the $\Lambda^{-n}$ factor) and so should all contribute
to Kondo screening.

This analysis suggests that in a situation with two electrons on the double dots which are bound into
a triplet by a putative RKKY coupling, there are nonetheless (at least)
three available screening channels, at least in the NRG reduction
of the Anderson Hamiltonian.  And so the problem would seem to be not one of an underscreened Kondo effect
but an exactly screened Kondo effect.  We do however note that while viewing the double dot system with
two dot electrons as an exactly screened Kondo effect is consistent with our Fermi liquid findings,
it is not clear under what conditions it would be correct to think of the two electron on the dots as ever forming
a triplet.  While any perturbatively generated RKKY interaction will generically be larger than the exponentially
small Kondo screening scale, there is a question of whether the zero temperature perturbation theory
underlying any RKKY estimate is convergent because the system is gapless.\cite{fye}

While we have focused here on NRG treatments of double dots close to their particle-hole symmetric point, similar
analyses of any multi-dot system might suggest that whenever the dot degrees of freedom exceed $S=1/2$,
it is not possible to ignore the $p\neq 0$ modes of the logarithmic basis.

\vskip .5in

Acknowledgements: MK was supported by the NSF under grant no. DMR-0906866.
RMK acknowledges support by the US DOE under contract number DE-AC02-98 CH 10886.  We thank Chung-Hou Chung, Alexei Tsvelik, Piers Coleman and Leonid Glazman for useful discussions.

\appendix

\section{Analysis of  the Ground State Entropy via the TBA Equations}

Here it is demonstrated that the ground state entropy of the double dot
system at zero temperature is zero and thus the ground state is a singlet.
The procedure outlined below can be found for a single dot in Section 8.3.3 of Ref. \onlinecite{wie}.

We start with the observation that the free energy of the system can be expressed as sums
over all excitations in the system, that is, over all possible solutions of the Bethe
ansatz equations (see Eqn. 10 of Ref. \onlinecite{long}).  Specifically it takes the form
\begin{eqnarray}
\Omega &=& E - TS,
\end{eqnarray}
where the energy of the system equals
\begin{eqnarray}
E &=& \int dk \rho(k) k + \sum^\infty_{n=1} \int d\lambda \sigma'_n (\lambda) \epsilon_{0n}(\lambda),
\end{eqnarray}
where
\begin{eqnarray}
\epsilon_{0n}(\lambda) &=& -n(2\epsilon_{d1}+U_1/2) + 2\int^\infty_{-\infty} a_n(\lambda-g(k))g(k);\cr\cr
a_n(x) &=& \frac{2}{\pi}\frac{1}{n^2+4x^2};\cr\cr
g(k) &=& \frac{(k-\epsilon_{d1}-\frac{U_1}{2})^2}{2U_1\Gamma_1},
\end{eqnarray}
and the entropy, $S$, is given by
\begin{eqnarray}
S &=& \int dk \bigg[(\rho(k) + \tilde\rho(k))\log (\rho(k) + \tilde\rho(k)) - \rho(k) \log\rho(k) -
\tilde\rho(k) \log (\rho(k))\bigg]\cr\cr
&& \hskip -.4in +
\sum^\infty_{n=0}\bigg[(\sigma_n(\lambda) + \tilde\sigma_n(\lambda))\log
(\sigma_n(\lambda ) + \tilde\sigma_n(\lambda )) - \sigma_n(\lambda) \log\sigma_n(\lambda) -
\tilde\sigma_n(\lambda) \log (\sigma_n(\lambda))\bigg]\cr\cr
&& \hskip -.4in +
\sum^\infty_{n=0}\bigg[(\sigma_n'(\lambda) + \tilde\sigma_n'(\lambda))\log
(\sigma_n'(\lambda ) + \tilde\sigma_n'(\lambda )) - \sigma_n(\lambda) \log\sigma_n'(\lambda) -
\tilde\sigma_n'(\lambda) \log (\sigma_n'(\lambda))\bigg].
\end{eqnarray}
Here $\rho(k),\sigma_n(\lambda),$ and $\sigma_n'(\lambda)$ are the particle densities
while $\tilde\rho(k),\tilde\sigma_n(\lambda),$ and $\tilde\sigma_n'(\lambda)$ are the hole
densities of the various excitations (i.e. solutions of the Bethe ansatz equations).
The particle and hole densities can be shown to obey the following equations:
\begin{eqnarray}
\tilde\rho(k) + \rho(k) &=& \rho_0(k)
- g'(k)\int d\lambda s(\lambda-g(k))(\tilde\sigma_1(\lambda)+\tilde\sigma_1'(\lambda)),
\end{eqnarray}
where
\begin{eqnarray}
\rho_0(k) &=& \frac{1}{2\pi} + \frac{1}{L}\Delta (k)
+ g'(k)\int d\lambda s(\lambda-g(k))(-\frac{1}{2\pi}\epsilon_{01}'(\lambda)
+ \frac{1}{L}\tilde\Delta_1(\lambda));\cr\cr
\Delta(k) &=& \frac{1}{2\pi}\partial_k \delta (k);\cr\cr
\tilde\Delta_n(\lambda) &=& -\frac{1}{\pi}Re\Delta(\sqrt{2U_1\Gamma_1}(\lambda + \frac{in}{2}));\cr\cr
s(x) &=& \frac{1}{2\cosh(\pi x)},
\end{eqnarray}
and
\begin{eqnarray}
\tilde\sigma_n(\lambda) + \sigma_n(\lambda) &=&
\int d\lambda' s(\lambda-\lambda'))(\tilde\sigma_{n+1}(\lambda')+\tilde\sigma_{n-1}(\lambda'))
+ \delta_{n1}\int dk \rho(k) s(\lambda-g(k));\cr\cr
\tilde\sigma_n'(\lambda) + \sigma_n'(\lambda) &=&
\int d\lambda' s(\lambda-\lambda'))(\tilde\sigma_{n+1}'(\lambda')+\tilde\sigma_{n-1}'(\lambda'))
+ D_n(\lambda),
\end{eqnarray}
where
\begin{eqnarray}
D_n(\lambda) &=& \delta_{n1}\int dk \rho(k) s(\lambda-g(k))
- \int d\lambda' s(\lambda-\lambda'))(\tilde\Delta_{n+1}(\lambda')+\tilde\Delta_{n-1}(\lambda'));
\end{eqnarray}
One sees that the density equations have source terms that involve a bulk piece and a piece
scaling as $1/L$, where $L$ is the system size.

One defines the energies of the excitations at finite temperature via the relations,
\begin{equation}
\epsilon (k) = T\log(\frac{\tilde\rho(k)}{\rho(k)});~~~
\epsilon_n (\lambda) = T\log(\frac{\tilde\sigma_n(\lambda)}{\sigma_n(\lambda)});~~~
\epsilon_n' (\lambda) = T\log(\frac{\tilde\sigma_n'(\lambda)}{\sigma_n'(\lambda)}).~~~
\end{equation}
These energies are given by the relations
$\epsilon(k),\epsilon_n(\lambda)$, and $\epsilon_n'(\lambda)$ which are governed by the equations,
\begin{eqnarray}
\epsilon(k) &=& k + \int d\lambda \epsilon_{01}(\lambda)s(\lambda - g(k)) + T\int d\lambda s(\lambda-g(k))
\log\bigg(\frac{n(\epsilon_1(\lambda))}{n(\epsilon_1'(\lambda))}\bigg);\cr\cr
\epsilon_n(\lambda) &=& \delta_{n1}T\int dk g'(k) s(\lambda - g(k))\log (n(-\epsilon(k)))\cr\cr
&& \hskip 1.5in
- T\int d\lambda' s(\lambda-\lambda')\log(n(\epsilon_{n-1}'(\lambda))n(\epsilon_{n+1}'(\lambda)));\cr\cr
\epsilon_n'(\lambda) &=& \delta_{n1}T\int dk g'(k) s(\lambda - g(k))\log (n(\epsilon(k)))\cr\cr
&& \hskip 1.5in - T\int d\lambda' s(\lambda-\lambda')\log(n(\epsilon_{n-1}'(\lambda))n(\epsilon_{n+1}'(\lambda))),
\end{eqnarray}
where $n(x) = (1+\exp(x/T))^{-1}$ is the Fermi function.
The equations for the $\epsilon$'s and the bulk pieces of the densities (i.e. the pieces not scaling
as $1/L$) are the same as for a single level dot.  As noted in the manuscript,
the Bethe ansatz equations for the double
dots in parallel are identical to the single level dot up to the impurity scattering phase.

Substituting the energies and densities in the expression for the free energy, one can rewrite it
in a much more simple fashion:
\begin{eqnarray}
\Omega &=& E_{gs} + T\int dk  \rho_0 (k) \log (n(-\epsilon(k))) +
T\int d\lambda \sum_{n=0}^\infty \rho^n_{gs}(\lambda)\log(n(\epsilon_n'(\lambda)));\cr\cr
E_{gs} &=& \sum^\infty_{n=0}\int d\lambda \epsilon_{0n}(\lambda)\rho^n_{gs}(\lambda);\cr\cr
\rho^n_{gs}(\lambda ) &=& \delta_{n1}\int \frac{dk}{2\pi}s(\lambda-g(k)) + \frac{1}{L}D_n(\lambda).
\end{eqnarray}
Here $E_{gs}$ is the ground state energy of the system.

One now wants to consider how $\Omega$ behaves at $T \rightarrow 0$.  If one can show that the leading
order correction in $\Omega$ at low temperatures is $T^2$ then as $S = -\partial_T \Omega$
one will have shown that the entropy vanishes as $T \rightarrow 0$, and so the ground state of the system
is a singlet.

In order to see that $\Omega$ has no term linear in $T$, it is sufficient to consider the
zero temperature values of the energies, $\epsilon (k)$ and $\epsilon_n'(\lambda)$.  At
the particle-hole symmetric point, one has
\begin{eqnarray}
\epsilon (k,T=0) > 0 &,& {\rm for~all~k};\cr\cr
\epsilon_1'(\lambda,T=0) < 0 &,& {\rm for~all~\lambda};\cr\cr
\epsilon_n'(\lambda,T=0) = 0 &,& n>1, ~{\rm for~all~\lambda}.
\end{eqnarray}
If one substitutes these expressions into the expression for $\Omega$ and uses the fact that
$\int d\lambda \rho^n_{gs}(\lambda) = 0$, $n>1$, one sees that $\Omega = E_{gs} + {\cal O}(T^2)$.

Now if one is away from the particle-hole symmetric point, one has instead
\begin{eqnarray}
\epsilon (k,T=0) > 0 &,& {\rm for~all~k};\cr\cr
\epsilon_n'(\lambda,T=0) = n(\frac{U_1}{2}+ \epsilon_1) &,& {\rm for~all~\lambda}.
\end{eqnarray}
Now while $\epsilon_1'(\lambda)$ is neither solely positive nor solely negative at zero temperature,
its leading order finite temperature correction is (see Section 8.3.7 of Ref. \onlinecite{wie}),
\begin{equation}
\epsilon_1'(\lambda,T) = \epsilon_1'(\lambda,T=0) + {\cal O}(T^2).
\end{equation}
Substituting these forms of the energies into the expression for the free energy, one
again sees that there is no term in $\Omega$ that is linear in T.

\section{Derivation of the Conductance in SBMFT}

Here we present a derivation of the conductance in the general case
of asymmetrically coupled dots. To determine the conductance we solve
the one-particle Schr\"odinger equation of the SBMFT Hamiltonian, $H_{SBMFT}|\psi\rangle = E|\psi\rangle$ where
$|\psi\rangle$ equals
\begin{eqnarray}
|\psi & > & =\int_{-\infty}^{+\infty}dxg_{1}(x)c_{1}^{\dagger}(x)|0>+
\int_{-\infty}^{+\infty}dxg_{2}(x)c_{2}^{\dagger}(x)|0>\nonumber \\
 &  & +\epsilon_{1}d_{1}^{\dagger}|0>+\epsilon_{2}d_{2}^{\dagger}|0>\label{eq:wavefunc}.
\end{eqnarray}
This gives the following four equations:
\begin{eqnarray}
-i\partial_{x}g_{1}(x)+\epsilon_{1}\tilde{V}_{11}\delta(x)+\epsilon_{2}\tilde{V}_{12}\delta(x) & = & Eg_{1}(x);\label{eq:eg1}\\
-i\partial_{x}g_{2}(x)+\epsilon_{1}\tilde{V}_{21}\delta(x)+\epsilon_{2}\tilde{V}_{22}\delta(x) & = & Eg_{2}(x);\label{eq:eg2}\\
\left(\tilde{\epsilon}_{d_{1}}-E\right)\epsilon_{1}+\tilde{V}_{11}g_{1}(0)+\tilde{V}_{21}g_{2}(0) & = & 0;\label{eq:zero1}\\
\left(\tilde{\epsilon}_{d_{2}}-E\right)\epsilon_{2}+\tilde{V}_{22}g_{2}(0)+\tilde{V}_{12}g_{1}(0) & = & 0.\label{eq:zero2}\end{eqnarray}
We then take the functions $g_{1,2}(x)$ found in the one particle wavefunction $|\psi\rangle$ to be of the following form:
\begin{eqnarray}
g_{1}(x) & = & e^{iEx}\left(\theta(-x)+R_{11}\theta(x)\right);\label{eq:g1}\\
g_{2}(x) & = & e^{iEx}\theta(x)T_{12}.\label{eq:g2}
\end{eqnarray}
Substituting the ansatz (Eqns. \ref{eq:g1} and Eq. \ref{eq:g2}) into
the above four equations,
we obtain four equations from which one can solve
for $T_{12}$. The conductance G, equal to $G=\frac{2e^2}{h}T_{12}T_{12}^{*}$,
is then
\begin{eqnarray}
G &=& \frac{2e^2}{h}\frac{D}{N}\cr\cr
N &=& 16\left[\tilde{V}_{12}\tilde{V}_{22}\tilde{\epsilon}_{d_{1}}+\tilde{V}_{11}\tilde{V}_{21}\tilde{\epsilon}_{d_{2}}\right]^{2}\cr\cr
D &=& 8\left[\tilde{V}_{11}\tilde{V}_{12}+\tilde{V}_{21}\tilde{V}_{22}\right]^{2}\tilde{\epsilon}_{d_{1}}\tilde{\epsilon}_{d_{2}}+16\tilde{\epsilon}_{d_{1}}^{2}\tilde{\epsilon}_{d_{2}}^{2}+\left[\tilde{V}_{12}\tilde{V}_{21}-\tilde{V}_{11}\tilde{V}_{22}\right]^{4}\cr\cr
&& \hskip 1in + 4\tilde{\epsilon}_{d_{1}}^{2}\left[\tilde{V}_{12}^{2}+\tilde{V}_{22}^{2}\right]^{2}+4\tilde{\epsilon}_{d_{2}}^{2}\left[\tilde{V}_{11}^{2}+\tilde{V}_{21}^{2}\right]^{2}.
\end{eqnarray}

We have also computed the transmission amplitude using
Ref. \onlinecite{meir-wingreen}:
\begin{equation}
T=\mbox{Tr}\left\{ G^{a}\tilde{\Gamma}_{R}G^{r}\tilde{\Gamma}_{L}\right\}, \label{eq:trace}
\end{equation}
where $G^{a/r}$ are advanced and retarded Green's function matrix
and $\tilde{\Gamma}_{R}$ and $\tilde{\Gamma}_{L}$ are defined by
\begin{equation}
\tilde{\Gamma}_{R}=\left[\begin{array}{cc}
\tilde{V}_{21}^{2} & \tilde{V}_{21}\tilde{V}_{22}\\
\tilde{V}_{21}\tilde{V}_{22} & \tilde{V}_{22}^{2}\end{array}\right],\qquad\tilde{\Gamma}_{L}=\left[\begin{array}{cc}
\tilde{V}_{11}^{2} & \tilde{V}_{11}\tilde{V}_{12}\\
\tilde{V}_{11}\tilde{V}_{12} & \tilde{V}_{12}^{2}\end{array}\right].
\end{equation}
We find that this trace formula (Eq. \ref{eq:trace}) gives exactly
the same result.

In the symmetric case (i.e. $\epsilon_{d1}=\epsilon_{d2}\equiv\epsilon_d$),
the above mean field equations reduce to two equations given by:
\begin{eqnarray}
0 &=& \frac{\tilde{\Gamma}}{\Gamma}-1+\frac{1}{\pi}\tan^{-1}\left[\frac{2\tilde{\Gamma}}{\tilde{\epsilon}_{d}}\right];\cr\cr
0 &=& \frac{\tilde{\epsilon}_{d}-\epsilon_{d}}{\Gamma}+
\frac{1}{\pi}\log\left[\frac{\left(\tilde{\epsilon}_{d}^{2}+4\tilde{\Gamma}^{2}\right)}{D^{2}}\right].
\end{eqnarray}
Here one can compute $T_{k}$ by solving
$T_{k}=\epsilon_{d}-\frac{N\Gamma}{\pi}\log\left[\frac{T_{k}}{D}\right]$
which in the limit $T_{k}\ll|\epsilon_{d}|$ gives
$T_{k}=De^{\frac{\pi\epsilon_{d}}{N\Gamma}}$.
In this case the Kondo limit is reached when the gate voltage reaches $\epsilon_{d}=-4.45$
(with $\Gamma=1$ and $D=100$).

\end{document}